\documentclass{aa}  

\usepackage{graphicx}
\usepackage{txfonts}
\begin{document}

   \title{Substellar multiplicity in the Hyades cluster}


   \author{G. Duch\^ene
          \inst{1,2}
          \and
          J. Bouvier\inst{2}
	\and
	E. Moraux\inst{2}
	\and
	H. Bouy\inst{3}
	\and
	Q. Konopacky\inst{4}
	\and
	A. M. Ghez\inst{5}
          }

   \institute{Astronomy Department, University of California, Berkeley, CA 94720-3411 USA\\
              \email{gduchene@berkeley.edu}
         \and
             UJF-Grenoble 1/ CNRS-INSU, Institut de Plan\'etologie et d'Astrophysique de Grenoble (IPAG), UMR 5274, Grenoble F-38041, France
	\and
	Centro de Astrobiolog\'{\i}a, Instituto Nacional de T\'ecnica Aerospacial, 28850 Torr\'ejon de Ardoz, Madrid, Spain
	\and
	Dunlap Institute for Astronomy and Astrophysics, University of Toronto, Toronto, ON M5S 3H4 Canada
	\and
	UCLA Division of Astronomy \& Astrophysics, Los Angeles CA 90095-1562, USA
             }

   \date{Received ...; accepted ...}

 
  \abstract
   {}
   {We present the first high-angular resolution survey for multiple systems among very low-mass stars and brown dwarfs in the Hyades open cluster.}
   {Using the Keck\,II adaptive optics system, we observed a complete sample of 16 objects with estimated masses $\lesssim 0.1\,M_\odot$.}
   {We have identified three close binaries with projected separation $\lesssim 0$\farcs11, or $\lesssim 5$\,AU. A number of wide, mostly faint candidate companions are also detected in our images, most of which are revealed as unrelated background sources based on astrometric and/or photometric considerations. The derived multiplicity frequency, 19$^{+13}_{-6}$\% over the 2--350\,AU range, and the rarity of systems wider than 10\,AU are both consistent with observations of field very low-mass objects. In the limited 3--50\,AU separation range, the companion frequency is essentially constant from brown dwarfs to solar-type stars in the Hyades cluster, which is also in line with our current knowledge for field stars. Combining the  binaries discovered in this surveys with those already known in the Pleiades cluster reveals that very low-mass binaries in open clusters, as well as in star-forming regions, are skewed toward lower mass ratios ($0.6 \lesssim q \lesssim 0.8$) than are their field counterparts, a result that cannot be accounted for by selection effects. Although the possibility of severe systematic errors in model-based mass estimates for very low-mass stars cannot be completely excluded, it is unlikely to explain this difference.}
   {We speculate that this trend indicates that surveys among very low-mass field stars may have missed a substantial population of intermediate mass ratio systems, implying that these systems are more common and more diverse than previously thought.}


   \maketitle
%

\section{Introduction}

Multiplicity is a common, if not ubiquitous, property of stellar populations from brown dwarfs to high-mass stars \citep[][and references therein]{duchene13}. Its characteristics and dependencies on primary mass \citep[and/or system mass; see][]{goodwin13} inform us on the physical processes involved in the star formation process itself and in the internal and external dynamical perturbations that affect newly formed systems. Akin to the question of the universality of the initial mass function, it remains undecided whether all star-forming environments lead to identical sets of multiplicity properties, such as their frequency, orbital period and mass ratio distributions. The population of field objects in the solar neighborhood provides a unique opportunity to detect companions of all masses and at all separations, but it represents an average over all types of star-forming environments. Stellar clusters, on the other hand, represent uniform populations that enable studies across primary masses, although interpretations must consider the possible disruptive influence of close stellar encounters. For instance, it has been shown that the multiplicity properties of field stars, the 120\,Myr-old Pleiades open cluster, the 1\,Myr-old Orion Nebula Cluster and the 1\,Myr-old Taurus loose star-forming association can all be accounted for with a unique set of initial conditions and purely dynamical effects \citep{kroupa01, kroupa03, parker11, marks12}.

\begin{table*}
\caption{\label{tab:sample}Observed sample}
\centering
\begin{tabular}{lccccccc}
\hline\hline
Target & Obs. Date & $J$ & $H$ & $K$ & $M (M_\odot)$ & Ref. \\
\hline
RM\,231 & 2008 Dec. 18 & 11.12 & 10.56 & 10.25 & 0.306--{\bf 0.321} & 2 \\
RM\,346\tablefootmark{a} & 2008 Dec. 20 & 11.18 & 10.55 & 10.28 & 0.308--{\bf 0.317} & 2 \\
RM\,132 & 2008 Dec. 18 & 11.16 & 10.57 & 10.28 & 0.305--{\bf 0.317} & 2 \\
RM\,221\tablefootmark{a} & 2008 Oct. 20 & 11.19 & 10.59 & 10.32 & 0.302--{\bf 0.310} & 2 \\
RM\,49\tablefootmark{a} &  2008 Dec. 20 & 11.21 & 10.59 & 10.33 & 0.302--{\bf 0.309} & 2 \\
RM\,182 & 2008 Dec. 20 & 11.24 & 10.63 & 10.36 & 0.296--{\bf 0.304} & 2 \\
RM\,60 & 2008 Dec. 20 & 11.34 & 10.76 & 10.47 & 0.277--{\bf 0.287} & 2 \\
RM\,376 & 2008 Dec. 20. & 11.38 & 10.79 & 10.53 & 0.273--{\bf 0.280} & 2 \\
RM\,126\tablefootmark{b} & 2008 Oct. 20 & 12.06 & 11.47 & 11.16 & 0.192--{\bf 0.200} & 2 \\ 
CFHT\,13 & 2009 Dec. 8 & 12.45 & 11.87 & 11.51 & 0.160--{\bf 0.170} & 1 \\
\hline
RM\,165\tablefootmark{c} & 2009 Dec. 7 & 13.49 & 12.86 & 12.51 & 0.107--{\bf 0.111} & 2 \\
CFHT\,19 & 2008 Oct. 20 & 14.16 & 13.43 & 12.90 & 0.085--{\bf 0.099} & 1 \\
UKIDSS\,1 & 2008 Oct. 20 & 14.60 & 13.85 & 13.42 & 0.075--{\bf 0.084} & 3 \\
UKIDSS\,8 & 2009 Dec. 7 & 15.60 & 14.55 & 14.02 & 0.060--{\bf 0.070} & 3 \\
UKIDSS\,5 & 2009 Dec. 7 & 15.81 & 14.72 & 14.05 & 0.057--{\bf 0.069} & 3 \\
UKIDSS\,3\tablefootmark{c} & 2009 Dec. 7 & 15.75 & 14.78 & 14.17 & 0.058--{\bf 0.066} & 3 \\
UKIDSS\,4 & 2008 Oct. 20 & 15.60 & 14.97 & 14.23 & 0.059--{\bf 0.065} & 3 \\
UKIDSS\,6 & 2009 Dec. 7 & 15.50 & 14.81 & 14.25 & 0.061--{\bf 0.064} & 3 \\
UKIDSS\,2 & 2009 Dec. 8 & 15.94 & 14.81 & 14.26 & 0.056--{\bf 0.064} & 3 \\
UKIDSS\,11 & 2008 Oct. 20 & 16.11 & 15.05 & 14.28 & 0.054--{\bf 0.063} & 3 \\
UKIDSS\,7 & 2008 Oct. 20 & 15.99 & 15.03 & 14.36 & 0.055--{\bf 0.061} & 3 \\
UKIDSS\,9 & 2008 Oct. 20 & 16.68 & 15.29 & 14.52 & 0.045--{\bf 0.058} & 3 \\
UKIDSS\,12 & 2008 Oct. 20 & 16.73 & 15.77 & 14.80 & 0.044--{\bf 0.054} & 3 \\
UKIDSS\,10 & 2009 Dec. 7 & 16.54 & 15.43 & 14.84 & 0.049--{\bf 0.053} & 3 \\
CFHT\,20 & 2008 Oct. 20 & 17.02 & 16.51 & 16.08 & 0.037--{\bf 0.039} & 1 \\
CFHT\,21 & 2008 Oct. 20 & 18.48 & 17.36 & 16.59 & 0.020--{\bf 0.035} & 1 \\
\hline
\end{tabular}
\tablefoot{The horizontal line demarcates low-mass from very low-mass targets. Masses are estimated (using integrated system magnitudes for binary systems) comparing the near-infrared magnitude to BT-Settl models at an age of 600\,Myr. Boldfaced entries indicate the masses derived from the $K$ band photometry. References: 1) \cite{bouvier08}; 2) \cite{reid93}; 3) \cite{hogan08}.\\
\tablefoottext{a}{Known binary \citep{reid97}.}
\tablefoottext{b}{Suspected binary \citep{reid97}.}
\tablefoottext{c}{These objects were re-observed on 2010 Dec. 09.}
}
\end{table*}

The multiplicity properties of very low-mass stars and brown dwarfs are markedly different from those of higher mass stars \citep{luhman12, duchene13}. Among field objects, the overall multiplicity frequency is on the order of 20\%, well below that of solar-type stars. In addition, most very low-mass multiple systems are both close (separation $\leq$\,25\,AU) and near equal-mass ($q  = M_B / M_A \gtrsim 0.8$, where A and B denote the most and least massive components, respectively), which differs from both low-mass and solar-type multiple systems. Similar trends are observed in star-forming regions, although a small population of wide systems is observed out to separations of at least 300\,AU. Despite these differences, there is no strong evidence that multiple system formation proceeds through different paths in the substellar and stellar regimes, except for the fact that "unequal" and wide binaries are increasingly uncommon for the lowest mass primaries. 

At intermediate ages, surveys of brown dwarf binaries in the Pleiades open cluster have revealed only a handful of tight systems \citep{martin03, bouy06}, again in agreement with trends in other populations. However, because of the distance to the Pleiades (125\,pc), the projected separations of most very low-mass binaries is below the capabilities of even adaptive optics systems on large ground-based telescopes or the Hubble Space Telescope. To probe multiplicity in open clusters, the Hyades represent an ideal opportunity owing to its close distance to the Sun \citep[46.3\,pc,][]{perryman98}. High-resolution surveys for visual binaries have been conducted in the Hyades for low-mass, solar-type and intermediate-mass stars, providing a solid comparison basis \citep{mason93, reid97, patience98}. Until recently, however, no substellar object was known in the Hyades cluster. \cite{bouvier08} and \cite{hogan08} identified a population of L- and T-type substellar cluster members based on their photometric and proper motion properties. While these imaging surveys exclude the presence of companions at separations of 100\,AU or more, nothing is known about companions in the typical range for very low-mass binaries, 1--10\,AU, which can only be probed with high-angular resolution techniques.

In this work, we present the first multiplicity survey of substellar members of the Hyades cluster, conducted with the adaptive optics systems installed on the 10\,m Keck\,II telescope. Our sample definition and observations are described in Section\,\ref{sec:obs}. Section\,\ref{sec:results} presents all candidate companions detected in this survey and provides astrometric and photometric arguments to discriminate between bound companions and unrelated background sources. Finally, Section\,\ref{sec:discuss} places the results of this survey in the broader context of multiplicity as a function of primary mass and environment.


\section{Sample, observations and data reduction}
\label{sec:obs}

We constructed our sample based on wide-area surveys that identified very low-mass members of the Hyades cluster based on their photometry and proper motion \citep{bouvier08, hogan08}. Based on the 600\,Myr isochrone of the BT-Settl evolutionary models \citep{allard03, allard09}, we find that the substellar limit occurs at $I \approx 18$ and $K \approx 14$ for a nominal cluster distance of 46.3\,pc, which we adopt throughout this study in the absence of parallax measurements. There are 13 known cluster members below that limit and we observed all of them in this study. Very low-mass stellar objects ($M \lesssim 0.1\,M_\odot$) are usually grouped with brown dwarfs for multiplicity statistics and other physical studies as it is generally assumed that the formation process does not abruptly change at the substellar limit. We therefore also included observations of all 3 known very low-mass cluster members with $K \geq 12.5$ (UKIDSS\,1, CFHT\,19 and RM\,165), yielding a total sample of 16 very low-mass cluster members that is complete to the best of our knowledge. We note that UKIDSS\,1 may itself be substellar if it is located on the close side of the cluster, as suggested by \cite{hogan08}. This highlights the complication induced by the depth of the cluster, which represents a significant fraction of its distance \citep[on the order of 30--40\%,][]{perryman98}. In addition, we also observed 10 objects in the 0.17--0.32\,$M_\odot$ range, including four known or suspected binaries from Hubble Space Telescope imaging \citep{reid97}.

While our sample of very low-mass objects includes all known members with $K \geq 12.5$, we note that the integrated brightness of near equal-mass binaries whose components are only slightly fainter than this limit would be above our selection threshold. In the absence of a spectroscopy-based sample definition, our sample may thus be biased {\it against} multiple systems. However, this bias does not apply to substellar objects given the wide gap seen in $K$ band luminosity between stellar and substellar objects \citep{hogan08}. It is also possible that the closest cluster members have their mass overestimated by our use of the nominal cluster distance, although that bias applies to both single objects and multiple systems. On the other hand, our sample is unlikely to be affected by the traditional Malmquist bias since all targets identified by \cite{bouvier08} and \cite{hogan08} lie at least 1\,mag above the detection limits of these surveys. In particular, it is extremely unlikely that these surveys missed any L-type member of the cluster in the area they covered, since these lie several magnitude above the detection limits.

   \begin{figure}
   \centering
   \includegraphics[width=\hsize]{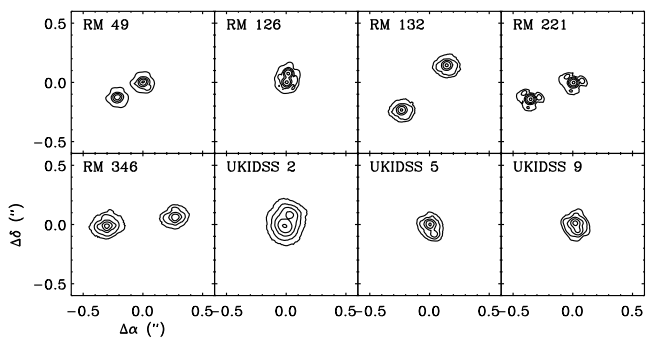}
      \caption{Contour plots of all subarcsecond binaries detected in this survey. All images were taken with the $K$ filter, except for UKIDSS\,2, which was observed with the $K'$ filter. In each panel, the highest contour is placed at factors $2^{-n}$ ($0 \leq n \leq 5$) times 90\% of the peak pixel.
              }
         \label{fig:bin}
   \end{figure}

The properties of all objects observed in this study are presented in Table\,\ref{tab:sample}. Target masses were estimated by comparing each object's near-infrared brightness to the 600\,Myr isochrones of the BT-Settl evolutionary models. The masses we derive from the objects' $K$ band brightness using the BT-Settl models differ only marginally ($\lesssim 5\%$) from those derived by \cite{bouvier08} using the NextGen \citep{baraffe98} and DUSTY \citep{chabrier00} models. Larger uncertainties are introduced by the use of brightness in multiple filters, however. The range quoted in Table\,\ref{tab:sample} for each target indicates the dispersion of estimates for a given object; to ensure uniformity, we adopt the masses derived from the $K$-band brightness throughout the paper. For tight binaries, masses indicated in the Table are estimated using integrated system magnitude to ensure uniformity; individual component masses are estimated in Section\,\ref{subsec:compmass}.

\begin{table*}
\caption{\label{tab:bin}Relative photometry and astrometry for all candidate companions}
\centering
\begin{tabular}{lccccccc}
\hline\hline
System & $\rho$ (\arcsec) & $PA$ (\degr) & $\Delta J$ (mag) & $\Delta H$ (mag) & $\Delta K$ (mag) & Note \\
\hline
{\bf RM\,346\,AB} & 0.575$\pm$0.001 & 277.3$\pm$0.1 & 0.42$\pm$0.03 & 0.39$\pm$0.03 & 0.38$\pm$0.02 & \\
{\bf RM\,132\,AB} & 0.533$\pm$0.001 & 135.0$\pm$0.1 & 0.04$\pm$0.03 & 0.04$\pm$0.03 & 0.06$\pm$0.02 & \\
{\bf RM\,49\,AB} & 0.251$\pm$0.001 & 120.7$\pm$0.1 & 0.27$\pm0.05$ & 0.25$\pm$0.03 & 0.23$\pm$0.03 & \\
{\bf RM\,221\,AB} & 0.387$\pm$0.001 & 111.1$\pm$0.2 & & & 0.01$\pm$0.02 & \\
{\bf RM\,126\,AB} & 0.074$\pm$0.001 & 350.7$\pm$0.8 & & & 0.02$\pm$0.02 &  \\ 
RM\,165\,AB & 4.595$\pm$0.009 & 296.2$\pm$0.1 & & & 4.45$\pm$0.03 & (2009) \\
& 4.686$\pm$0.009 & 296.1$\pm$0.1 & & 4.38$\pm$0.03 & & (2010) \\
RM\,165\,AC & 7.331$\pm$0.011 & 75.1$\pm$0.1 & & & 4.81$\pm$0.08 & (2009) \\
 & 7.233$\pm$0.014 & 74.9$\pm$0.1 & & 4.50$\pm$0.03 & & (2010) \\
{\bf UKIDSS\,2\,AB} & 0.110$\pm$0.005 & 335.8$\pm$0.2 & & & 0.70$\pm$0.15\tablefootmark{a} & \\
UKIDSS\,2\,AC\tablefootmark{b} & 5.141$\pm$0.010 & 27.3$\pm$0.1 & & & 1.33$\pm$0.03\tablefootmark{a} & \\
UKIDSS\,3\,AB & 1.902$\pm$0.004 & 162.2$\pm$0.1 & & & 5.16$\pm$0.05 & (2009) \\
 & 1.855$\pm$0.005 & 165.2$\pm$0.1 & & & 5.14$\pm$0.07\tablefootmark{a} & (2010) \\
UKIDSS\,4\,AB & 3.731$\pm$0.007 & 53.62$\pm$0.06 & & & 4.91$\pm$0.05 & \\
{\bf UKIDSS\,5\,AB} & 0.093$\pm$0.002 & 204.9$\pm$1.2 & 1.8$\pm$0.3 & 1.6$\pm$0.1 & 1.35$\pm$0.05 & \\
UKIDSS\,6\,AB & 2.653$\pm$0.005 & 237.0$\pm$0.1 & 6.0$\pm$0.2 & 6.0$\pm$0.1 & 5.61$\pm$0.06 & \\
UKIDSS\,7\,AB\tablefootmark{b} & 5.743$\pm$0.011 & 223.3$\pm$0.1 & & & 2.75$\pm$0.02 & \\
{\bf UKIDSS\,9\,AB} & 0.066$\pm$0.002 & 199.8$\pm$1.7 & & 1.4$\pm$0.1 & 1.08$\pm$0.05 & \\
UKIDSS\,9\,AC & 6.667$\pm$0.013 & 310.6$\pm$0.1 & & 4.46$\pm$0.05 & 4.81$\pm$0.03 & \\
UKIDSS\,10\,AB\tablefootmark{b} & 3.722$\pm$0.007 & 54.5$\pm$0.1 & & & -1.33$\pm$0.02 & \\
UKIDSS\,12\,AB & 2.251$\pm$0.005 & 90.4$\pm$0.1 & & 4.37$\pm$0.04 & 4.94$\pm$0.07 & \\
CFHT\,21\tablefootmark{c} & 3.634$\pm$0.008 & 132.4$\pm$0.1 & & & 0.87$\pm$0.02 & \\
\hline
\end{tabular}
\tablefoot{Boldfaced entries indicate systems that are considered physically bound.\\
\tablefoottext{a}{These flux ratios were measured with $K'$ filter.}
\tablefoottext{b}{Companion detected in UKIDSS images \citep{hogan08}. The relative astrometry and photometry presented here is from our adaptive optics images.}
\tablefoottext{c}{Companion detected in CFHT-IR images \citep{bouvier08}. The relative astrometry and photometry presented here is from our adaptive optics images.}
}
\end{table*}

Observations were conducted using the adaptive optics system on the 10\,m Keck\,II telescope in the "laser guide star" mode \citep{wizi06}. Targets that are bright enough in the visible ($R \lesssim 17.5$) were used as tip-tilt reference stars, while others were observed off-axis using an unrelated star within 60\arcsec\ as reference star. Data were taken with the NIRC2 near-infrared instrument (P.I.: K. Matthews) and its narrow camera mode with a pixel-scale of 0\farcs00996/pix \citep{ghez08}, providing a total field of $\approx$10\arcsec. The survey was conducted with the $K$ (2.20$\mu$m) and $K'$ (2.12$\mu$m) filter, but we also obtained follow-up $H$ (1.63$\mu$m) and $J$ (1.25$\mu$m) images for some systems to estimate the colors of candidate companions. Images were taken in a 3-position pattern. At each position, two or three separate images were recorded, each being the co-addition of 1 to 6 individual frames whose integration time was chosen to optimize signal to noise while avoiding saturation. Individual integration times ranged from 3 to 60\,sec. Two systems were re-observed after one year with a similar observing sequence to discriminate between physically related companions and background stars.

The data reduction process included the usual steps for near-infrared images. An image of the background emission was generated by medianing all images of a given target and subtracted from each individual images. The resulting images were flat-fielded, cosmetically cleaned and shift-and-added to produce the final images. In a few cases, the quality of the adaptive optics correction was fluctuating and/or modest, and we selected the highest-quality frames to produce final images. While this improved the resulting point spread function (PSF) at close separation ($\lesssim0$\farcs25), it negatively affected the contrast of our images at larger separations. We therefore also inspected the final images combining all individual frames to search for faint, distant companions to these targets. The median FWHM achieved in our final images is 0\farcs066, with a standard deviation of 0\farcs019, and the largest FWHM is 0\farcs110. 

   \begin{figure}
   \centering
   \includegraphics[width=\hsize]{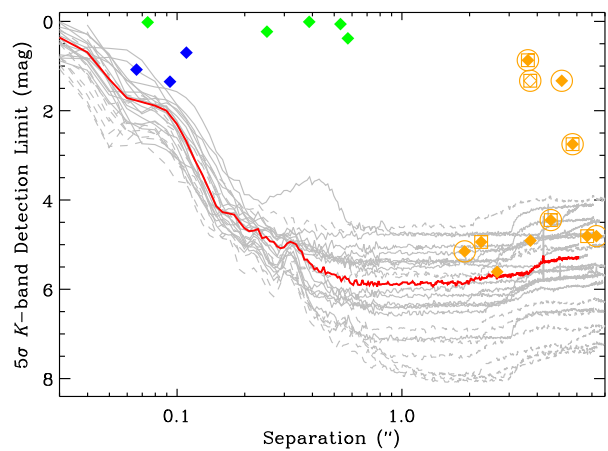}
      \caption{$K$- (or $K'$-) band detection limits ($5\sigma$) for companions. Individual curves are shown as gray curves, with solid (dashed) lines indicating very low mass (low mass) targets, respectively. The red curve shows the median detection limit for very-low mass targets. All curves are smoothed using a 3-pixel running median filter for separations larger than 0\farcs8 for visual purposes. Blue and green diamonds represent candidate companions to very low-mass and low-mass targets identified in our survey, respectively. Orange symbols indicate companions which are confirmed or suspected unrelated background stars based on astrometric (circle symbols) and/or photometric (square symbols) arguments. The magnitude difference to the bright visual companion to UKIDSS\,10 has been inverted and is shown as the open diamond.
              }
         \label{fig:lim}
   \end{figure}

The relative astrometry and photometry of all candidate binary systems was estimated in the following manner. For wide companions (projected separation $\ge$0\farcs25), the position of the companions were estimated using a Gaussian fit to the core of the PSF while the flux ratio was estimated using a small-radius aperture (typically 0\farcs05) and accounting for residual background emission. Uncertainties were estimated as the standard deviation of the mean based on all individual images containing the companions. Furthermore, we added in quadrature uncertainties of 0.2\% on the separation, 0\fdg1 on the position angle and 0.02\,mag on the flux ratio as estimates of the absolute precision of the method based on prior experience of such data. For tight binaries, we performed PSF fitting using the DAOPHOT package, using several single stars in our survey as candidate PSFs and adopting the one resulting in the smallest residuals. As a secondary check to account for the fact that none of these PSF is a perfect match, we also subtracted an azimuthally-symmetric PSF generated using a radial profile of the primary on the opposite of the companion to better highlight the companion. We conservatively adopt the half-range between PSF fitting and primary subtraction as an estimate of our final astrometric and photometric uncertainty although the latter method is likely to lead to significantly biased results. The final astrometric and photometric properties of all multiple systems detected in this survey are listed in Table\,\ref{tab:bin}.


\section{Results}
\label{sec:results}

   \begin{figure*}
\centering
            \includegraphics[width=4.5cm]{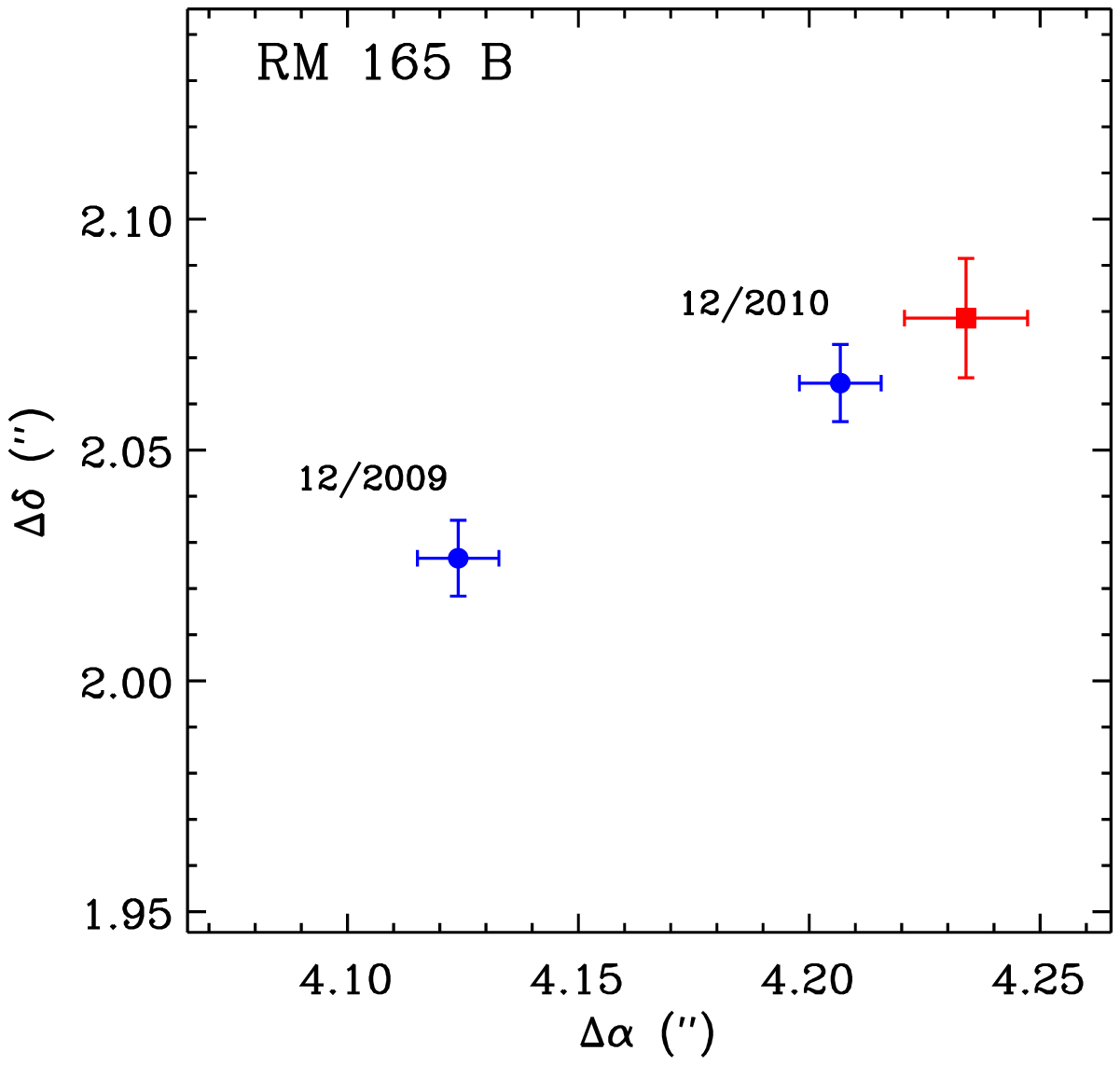}
            \includegraphics[width=4.5cm]{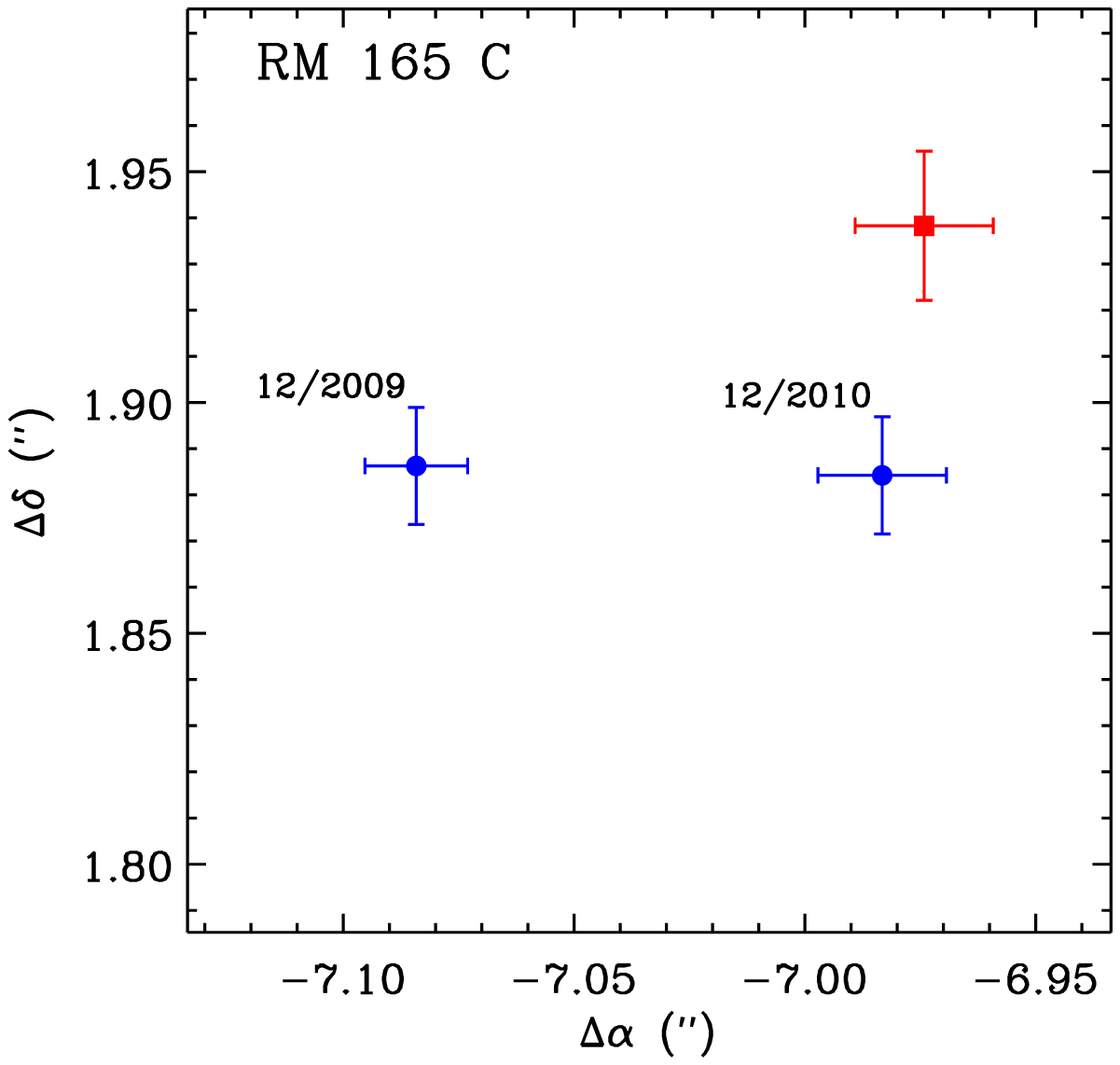}
            \includegraphics[width=4.5cm]{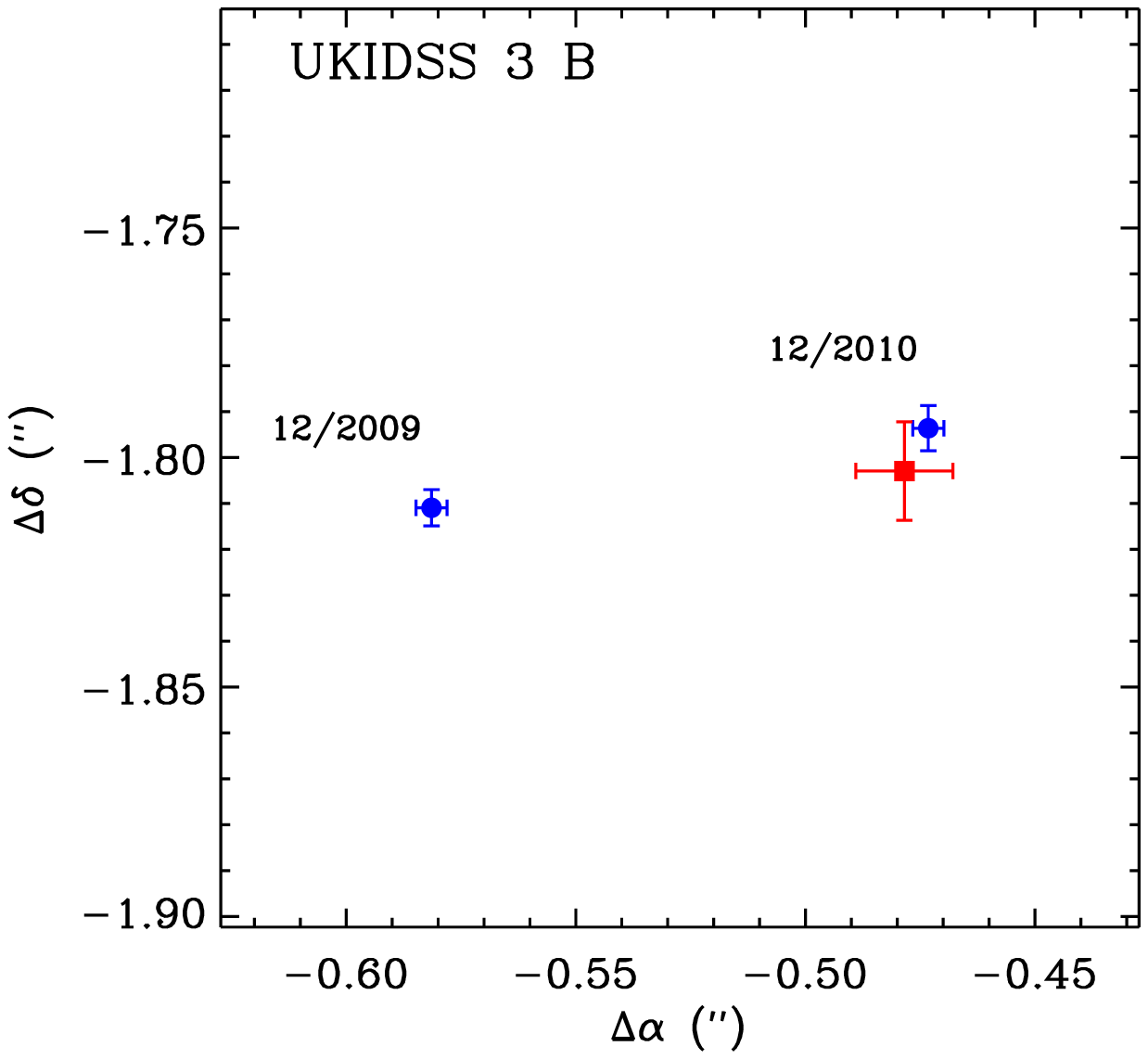}
            \includegraphics[width=4.5cm]{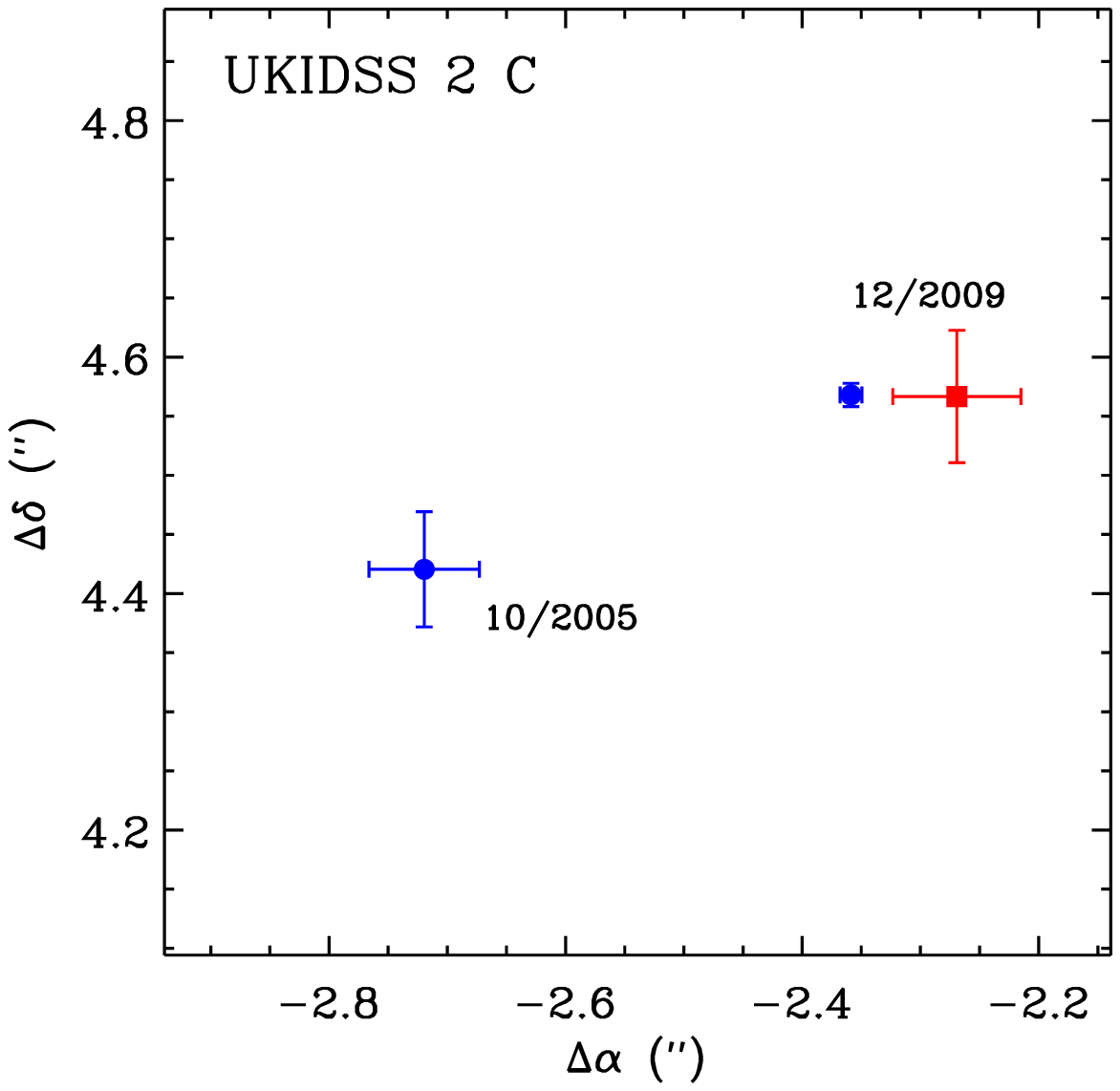}
            \includegraphics[width=4.5cm]{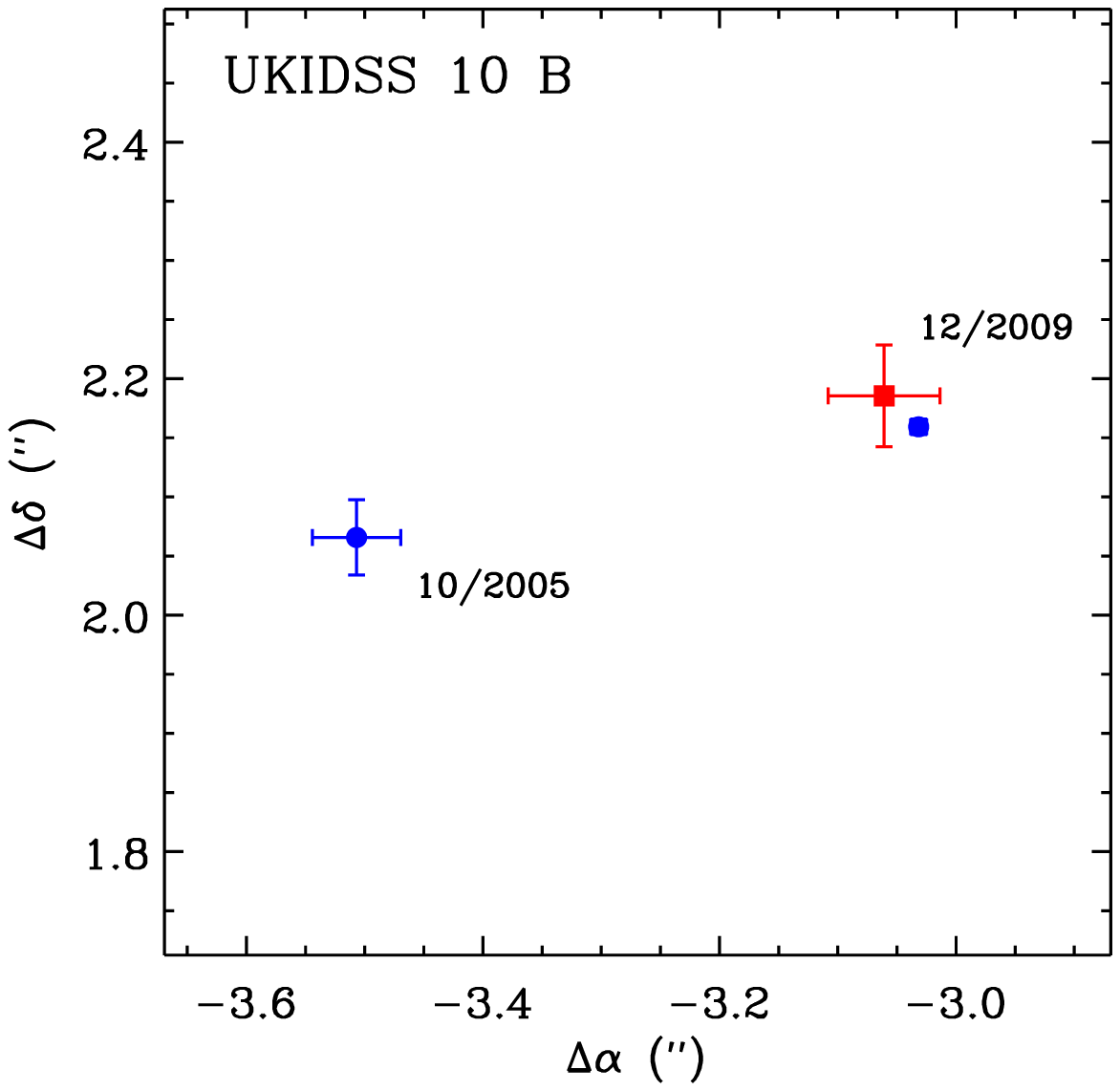}
            \includegraphics[width=4.5cm]{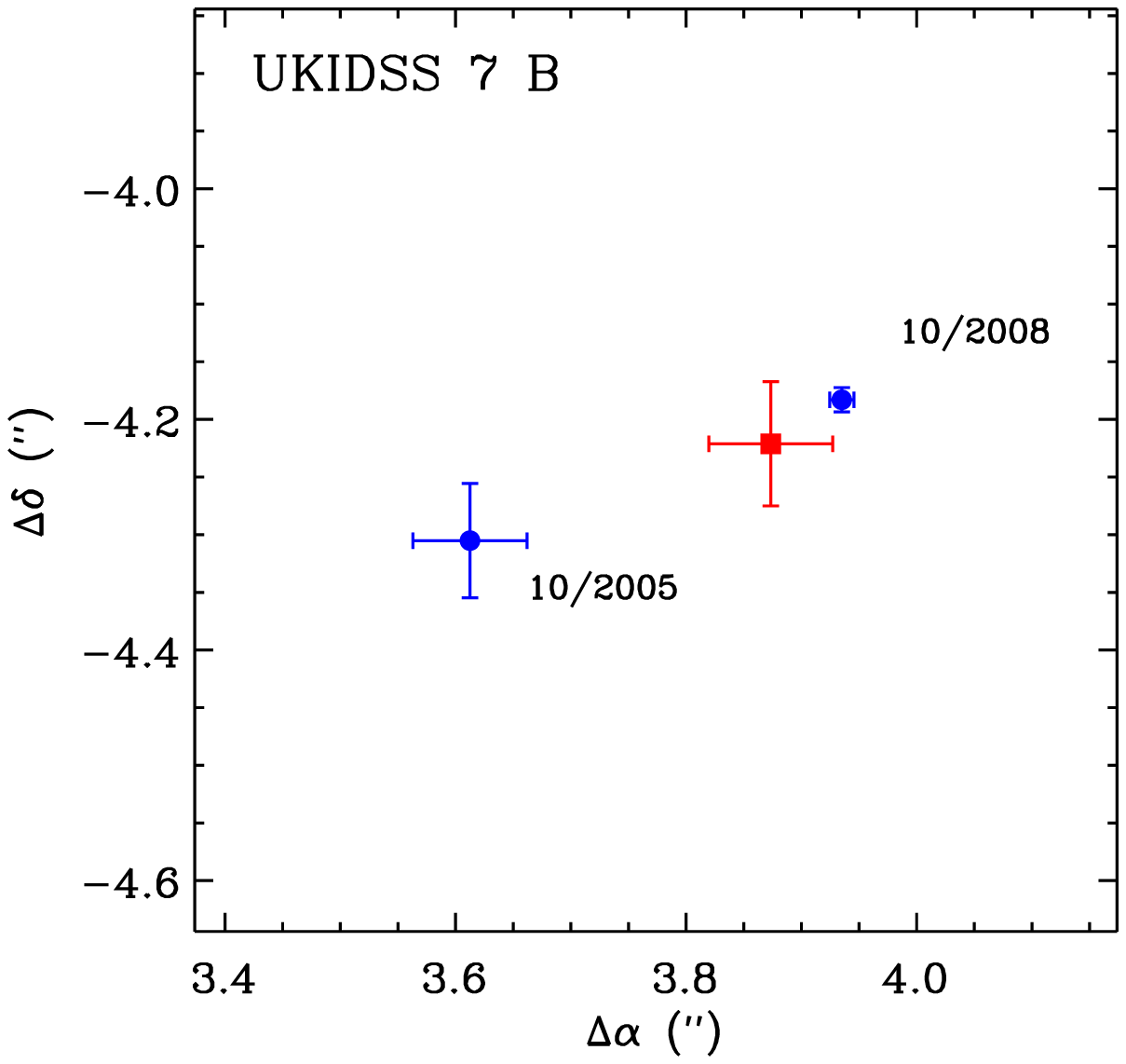}
            \includegraphics[width=4.5cm]{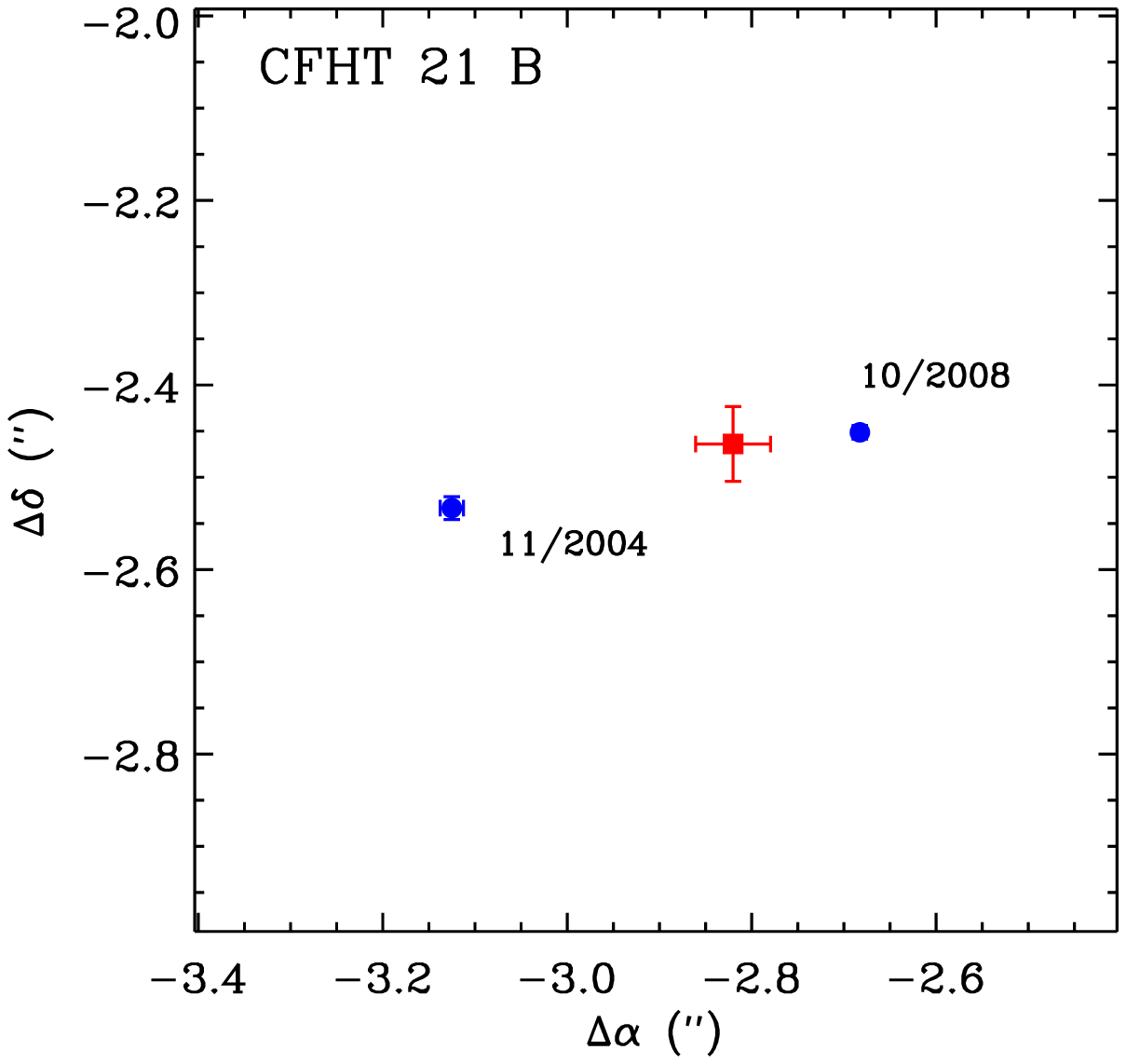}
      \caption{Observed relative motion for candidate companions observed at two epochs (blue circles). The red squares indicate the predicted position of the companion at the second epoch if it were a fixed background object based on the known proper motion of the primary (uncertainties compound proper motion uncertainty and first epoch position uncertainty). The first three panels present systems with adaptive optics imaging at two epochs while the last four make use of UKIDSS (UKIDSS\,2, 7 and 10), or CFHT12K (CFHT\,21) data as first epoch. Given the observing dates involved in this analysis, parallactic motion is negligible in these comparisons.}
         \label{fig:pm}
   \end{figure*}

\subsection{Detected candidate companions}
\label{subsec:comps}

Given the resolution of our images and our adopted observing sequence, we are able to find companions as close as $\approx$0\farcs05 (2.3\,AU at the distance of the cluster) and as wide as 7\farcs5 (and up to 10\arcsec\ in certain directions), approximately. Visual inspection of all images revealed that we detected 8 subarcsecond companions (shown in Figure\,\ref{fig:bin}), including the four known or suspected binaries first reported by \cite{reid97}. In addition, we found 11 companions at $\ge$1\farcs5 of their primaries. All companions are shown in Figure\,\ref{fig:lim}, along with our individual 5$\sigma$ detection limits which were estimated using the standard deviation in concentric annuli centered on the primary star and masking out any companion. The slight loss in sensitivity seen at $\approx 3$\arcsec\ in most curves is a consequence of our dithering pattern while the bump at $\approx 0$\farcs4 in the shallowest curve represents an artefact introduced by an improperly corrected "waffle" mode in the images of UKIDSS\,2. We further computed a median detection limit curve considering only the 16 very low-mass targets in our sample which constitute the core of our survey. Between 0\farcs5 and 4\arcsec, our median detection limit is $\Delta K \approx 6$\,mag, with a scatter of about 0.5\,mag.

Inspection of Figure\,\ref{fig:lim} suggests a separation between two categories of companions: subarcsec companions, all found to have $\Delta K \le 2$\,mag, on one hand, and wide companions, with $\Delta K$ as large as 6\,mag, on the other. While it is tempting to associate these categories with physical and non-physical companions, respectively, this can only be confirmed with additional information, such as proper motion or colors. In the following we try to make this discrimination for each system. We first consider the relative astrometry of all candidate binaries as this is the most powerful criterion to establish whether a system is physically bound.

\subsection{Astrometry of candidate companions}
\label{subsec:astrom}

First of all, let us consider RM\,49, 126, 221 and 346, whose binary status was known or suspected from previous high-resolution  imaging \citep{reid97} and is confirmed by our observations. While significant relative motion is observed in all four cases (20--30\% in projected separation and up to about 30\degr\ in position angle), its amplitude corresponds to an angular velocity of $\lesssim 20$\,mas/yr, much smaller than the proper motion of Hyades members \citep[$\gtrsim 100$\,mas/yr,][]{reid92}, thereby excluding that the companions are unrelated background stars. The observed relative motion of all four pairs is $\lesssim$2.5\,km/s, which is consistent with orbital motion for systems of total mass 0.1--0.2\,$M_\odot$ and semi-major axis 15--25\,AU. While the data at hand are too limited to estimate a dynamical mass for these systems, we note that the tightest of these systems is the one showing the least orbital motion, suggesting that projection effects are important for this system. We consider all four systems as physically bound. None of the other tight systems has prior high-resolution observations, including the low-mass system RM\,132, so no relative motion can be estimated. However, given their small angular separations and the relative scarcity of stars of similar brightness within our field of view, we estimate that all four systems are very likely physical bound.

\begin{table*}
\caption{\label{tab:pm}Relative motion within candidate wide binaries}
\centering
\begin{tabular}{lcccccc}
\hline\hline
System & $\Delta t$ (yr)\tablefootmark{a} & $\mu_\alpha$ & $\mu_\delta$ & $\mu_\alpha$ & $\mu_\delta$ & Ref. \\
 & & \multicolumn{2}{c}{Internal motion\tablefootmark{b}} & \multicolumn{2}{c}{Absolute motion\tablefootmark{c}} & \\
\hline
RM\,165\,AB & 1.0 & 83$\pm$12 & -38$\pm$12 & 110$\pm$10 & -52$\pm$10 & 1 \\
RM\,165\,AC & 1.0& 101$\pm$18 & 0$\pm$18 & 110$\pm$10 & -52$\pm$10 & 1  \\
RM\,165 (avg.) & 1.0 & 92$\pm$11 & -19$\pm$11 & 110$\pm$10 & -52$\pm$10 & 1  \\
UKIDSS\,2AC & 4.0 & 91$\pm$12 & -37$\pm$13 & 114$\pm$7 & -37$\pm$7 & 2  \\
UKIDSS\,3 & 1.0 & 108$\pm$5 & -17$\pm$6 & 102$\pm$7 & -8$\pm$7 & 2 \\
UKIDDS\,7 & 3.0 & 108$\pm$17 & -41$\pm$17 & 87$\pm$7 & -28$\pm$7 & 2 \\
UKIDSS\,10 & 4.1 & 115$\pm$9 & -23$\pm$8 & 108$\pm$7 & -29$\pm$7 & 2  \\
CFHT\,21 & 3.9 & 115$\pm$4 & -21$\pm$4 & 79$\pm$10 & -18$\pm$10 & 3 \\
\hline
\end{tabular}
\tablefoot{All proper motions are presented in mas/yr. References: 1) \cite{reid93}; 2) \cite{hogan08}; 3) \cite{bouvier08}.\\
\tablefoottext{a}{Time baselines of 1\,yr correspond to our own multi-epoch adaptive optics images, while the longer time baselines compare our adaptive optics images with UKIDSS or CFHT-IR images (see Section\,\ref{subsec:astrom}).}
\tablefoottext{b}{Relative motion of the primary component relative to the secondary.}
\tablefoottext{c}{Proper motion of the primary component measured in their discovery studies.}
}
\end{table*}

Three of the "wide" companions have been observed at two epochs in our study (RM\,165\,B and C, UKIDSS\,3\,B). Within one year, all of them appear to have moved by $\approx$0\farcs1 relative to their primary, which is much larger than both our astrometric uncertainties and their expected orbital motion given their wide separation. On the contrary, the observed motion matches within $\approx2\sigma$ the known proper motion of our targets, in both amplitude and direction, as shown in Figure\,\ref{fig:pm} and Table\,\ref{tab:pm}. Therefore, these "companions" are instead fixed background sources. In addition, the UKIDSS\,2C, UKIDSS\,7\,B and UKIDSS\,10C candidate companions were also detected by the original images of the UKIDSS survey \citep{hogan08} with sufficient signal-to-noise to precisely measure their position. As Figure\,\ref{fig:pm} and Table\,\ref{tab:pm} show, all three of these companions are also found to be unrelated background stars as their relative motion is much more consistent with the target's proper motions than with the hypothesis of a comoving object. Similarly, we reach the same conclusion for the apparent companion to CFHT\,21, which was also detected in the initial visible and near-infrared images obtained by \cite{bouvier08} with CFHT12K and CFHT-IR, respectively. Both the UKIDSS and CFHT-IR images provide a 3--4\,yr time baseline. For all seven candidate companions discussed here, we achieved a precision in the relative motion within the pairs of 20\,mas/yr or better.

\subsection{Photometry of candidate companions}
\label{subsec:photom}

\begin{figure}
   \centering
   \includegraphics[width=\hsize]{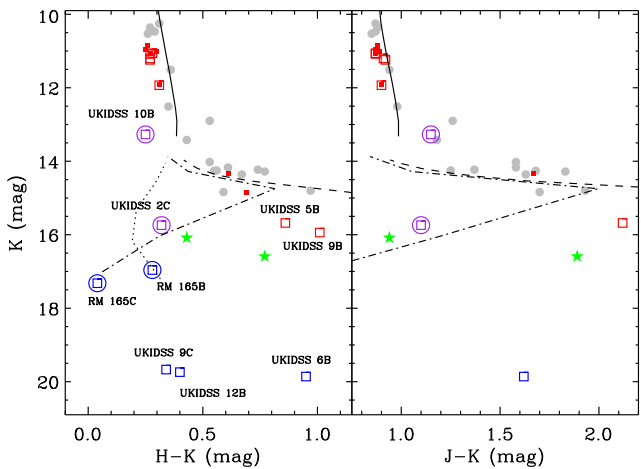}
      \caption{Color-magnitude diagrams for very low-mass Hyades members. Single M and L stars in our survey are shown as gray circles while the two T dwarfs are shown at green stars. Tight binaries ($\lesssim 0$\farcs5) are shown as pairs of filled and open red squares for primaries and secondaries, respectively. Wide companions detected in multiple bands in our adaptive optics images and in 2MASS images (in which case we use 2MASS photometry) are shown as blue and purple open squares, respectively. All candidate companions to very low-mass cluster members are labeled. Circled symbols indicate companions shown to be unrelated background stars by their motion relative to their respective primaries. The solid line represents the 600\,Myr NextGen isochrone down to a mass of 0.075\,$M_\odot$ while the dashed, dotted and dot-dashed curves represent the DUSTY, COND and BT-Settl models over the 0.03--0.07\,$M_\odot$ range. Note that the COND models predict $J-K$ colors that are markedly bluer than any known very-low mass cluster member.
              }
         \label{fig:cmd}
   \end{figure}

In addition to proper motion, colors can be used to evaluate whether companions are physically bound as all cluster members (primaries, companions or single stars) must define a clear sequence based on the cluster's age and absence of significant foreground extinction. Five wide companions in our survey were detected at more than one wavelength in this survey (Figure\,\ref{fig:cmd}). Three of these (RM\,165\,C, UKIDSS\,9C and UKIDSS\,12B) are both markedly bluer ($\Delta (H-K) \approx 0.4$\,mag) and much fainter ($\Delta K \approx4.8$\,mag) than their respective primaries. While such colors could in principle be explained if the companions were T dwarfs, their faintness (e.g., relative to CFHT\,20) makes this interpretation unlikely. RM\,165\,C, which is less than 1\,mag fainter than CFHT,21, has been shown to be a background star (see Section\,\ref{subsec:astrom}). By extension, we consider that this is also true for UKIDSS\,9C and UKIDSS\,12B. Of the remaining two candidate companions with multi-color data, RM\,165\,B has been excluded by its (lack of) apparent proper motion. While the near-infrared colors of UKIDSS\,6\,B are similar to those of its primary, its extreme faintness ($K \approx 19.9$) strongly suggests that it also is a background star. Finally, a handful of our targets and candidate companions have been detected at optical wavelengths and can be found in the USNO-B and SDSS catalogs or have been observed by \cite{bouvier08}. Specifically, we find extremely red color for the bona fide cool Hyades members RM\,165\,A and UKIDSS\,7\,A ($R-K\approx6.0$ and 8.7\,mag, respectively), and comparatively much bluer colors for distant candidate companions ($R-K\approx4.2$ and 3.4\,mag for UKIDSS\,7B and 10B, respectively). Since UKIDSS\,10B has already failed the common proper motion test, we conclude that UKIDSS\,7B is also a background star. Similarly, CFHT\,21B is much bluer ($I-K=2.5$ vs $I-K=5.6$), and even brighter in the $I$ and $z$ bands, than CFHT\,21A \citep{bouvier08}, clearly indicating that it is not an extremely low-mass companion. 

In summary, we conclude that all subarcsecond binary found or confirmed in our survey are physically bound companions. On the other hand, most companions at larger projected separations are identified unrelated background stars based on proper motion analysis (for 7 out of 11 objects) and/or their near-infrared colors. The only 2 wide companions for which neither of these criteria are available or conclusive are UKIDSS\,4\,B and UKIDSS\,6\,B. Both are much fainter than any known cluster members ($K \geq 19$), and would have masses in the 10--20\,$M_{Jup}$ range if they were cluster members. While it cannot be excluded that these are indeed extremely low-mass companions on wide orbits, their brightness is in the same range as those of other wide companions that were excluded in the previous analysis. We also note that the near-infrared color of UKISS\,6\,B are consistent with those of L dwarfs or extincted M dwarfs, but are inconsistent with a T dwarf spectral type that would be expected for such low mass objects. We therefore conclude that these two companions are most likely unrelated background star although follow-up observations are warranted to confirm this. 

\subsection{Estimated component masses}
\label{subsec:compmass}

For each tight (subarcsecond) binary in our survey we estimate individual component masses by comparing their near-infrared brightness to the 600\,Myr isochrone of the BT-Settl evolutionary models (see Table\,\ref{tab:mass}). As can be seen in Figure\,\ref{fig:cmd}, the evolutionary models are not a particularly good fit to the near-infrared photometry of our substellar targets, especially for faintest objects ($K \gtrsim 15$), implying that the derived masses should be considered with caution. We also considered the AMES-COND and DUSTY models \cite{chabrier00}, although neither produces a satisfying isochrone either. In the 0.05--0.10\,$M_\odot$, typical differences between these various models are on the order of 2\% for stellar objects and 5--15\% for objects with $K \lesssim 16$ ($M \gtrsim 0.04\,M_\odot$). In particular, we note that the uncertainties introduced by the difficulty in measuring tight, unequal-flux binaries are generally smaller than those inherent to our use of a particular evolutionary model.

As could be expected, the masses for the individual components of the stellar binary systems are revised downwards by about 30\% relative to the initially estimated system mass. On the other hand, the primaries of the other tight binary systems (UKIDSS\,2, 5 and 9) uncovered in this survey have estimated masses that are only $\sim15$\% lower than those estimated from the integrated system magnitudes as a result of their unequal near-infrared flux ratios. Due to the steep mass-luminosity relationship predicted by the evolutionary models, their companions are only somewhat less massive: all three have similar masses of 0.040--0.046\,$M_\odot$. Indeed, the brightness of these components is intermediate between the L dwarfs objects discovered by \cite{hogan08} and the faintest known (T-type) Hyades members from \cite{bouvier08}. Photometric and spectroscopic follow-up is required to obtained more precise mass estimates for these objects.

\begin{table}
\caption{\label{tab:mass}Estimated masses for components of tight Hyades binaries}
\centering
\begin{tabular}{lcccccc}
\hline\hline
Object & $M_J (M_\odot)$ & $M_H (M_\odot)$ & $M_K (M_\odot)$ & $q$\\
\hline
RM\,49\,A & 0.220 & 0.219 & 0.222 & \\
RM\,49\,B & 0.191 & 0.192 & 0.196 & 0.87--{\bf 0.88} \\
RM\,132\,A & 0.215 & 0.210 & 0.218 & \\
RM\,132\,B & 0.210 & 0.205 & 0.211 & {\bf 0.97}--0.98 \\
RM\,346\,A & 0.232 & 0.230 & 0.234 & \\
RM\,346\,B & 0.187 & 0.189 & 0.192 & 0.81--{\bf 0.82} \\
RM\,221\,A & 0.209 & 0.205 & 0.210 & \\
RM\,221\,B & 0.208 & 0.204 & 0.209 & {\bf 0.99} \\
RM\,126\,A & 0.139 & 0.137 & 0.141 & \\
RM\,126\,B & 0.138 & 0.136 & 0.140 & {\bf 0.99} \\
UKIDSS\,2\,A & & & 0.055 & \\
UKIDSS\,2\,B & & & 0.046 & {\bf 0.84} \\
UKIDSS\,5\,A & 0.055 & 0.059 & 0.062 & \\
UKIDSS\,5\,B & 0.030 & 0.037 & 0.043 & 0.55--{\bf 0.69} \\
UKIDSS\,9\,A & & 0.052 & 0.053 & \\
UKIDSS\,9\,B & & 0.033 & 0.040 & 0.63--{\bf 0.75} \\
\hline
\end{tabular}
\tablefoot{Masses are estimated by comparing each object's brightness in a given filter with the 600\,Myr-old BT-Settl isochrone \citep{allard03, allard09} using a distance modulus of 3.33 and assuming no foreground extinction. Boldfaced entries indicate mass ratios derived from the $K$ band photometry of the individual components.
}
\end{table}

\begin{figure}
   \centering
   \includegraphics[width=\hsize]{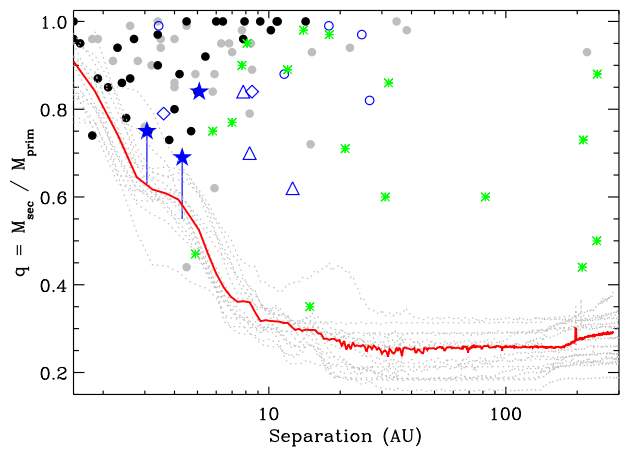}
      \caption{Detection limits ($5\sigma$) for companions to Hyades very low-mass stars expressed as a minimum detectable mass ratio derived from the 600\,Myr $K$-band BT-Settl isochrone. Dotted gray curves represent individual detection limits while the red curve shows our median detection limit. Blue stars represent companions to Hyades brown dwarfs with mass ratios estimated from their $K$ band photometry; the associated vertical segments indicate the range of mass ratios derived from the various available near-infrared fluxes. Only companions believed to be bound to our targets are shown here. Blue diamonds represent Hyades very low-mass binaries from \cite{siegler03}, for which we recomputed component masses using the same method as systems from our survey. Open blue triangles represent Pleiades brown dwarf binaries from \cite{martin03} while green asterisks represent binary very low-mass objects in nearby star forming regions (age $\lesssim 5$\,Myr, see Section\,\ref{subsec:massratio}). Blue circles indicate the low-mass binaries observed in this study. Gray and black circles represent field very low-mass stellar and substellar binaries from the Very Low-Mass Binary archive, respectively. Notice how brown dwarf binaries in open clusters and star-forming regions tend to have lower mass ratios relative to their field counterparts.
              }
         \label{fig:qlim2}
   \end{figure}

Finally, based on the $K$-band 600\,Myr isochrone, we note that we could have detected companions down to $q_{min} \approx 0.3$ at physical separations of 10\,AU or more, as illustrated in Figure\,\ref{fig:qlim2}. In the 3--5\,AU range where companions to very low-mass stars are found, however, our observational detection limit corresponds to a minimum detectable mass ratio of $q_{min} \approx 0.6$. At a separation of 2\,AU ($\approx$0\farcs045), which we consider the limit of our survey, our observations are sensitive to companions with $q \gtrsim 0.8$.

\section{Discussion}
\label{sec:discuss}

\subsection{Multiplicity as a function of mass}

We first discuss the multiplicity frequency of very low-mass stars in the Hyades cluster. We only consider our sample of such objects and do not include the two candidate very low-mass binaries discovered by \cite{siegler03}. For one, we do not know which other Hyades members they targeted in their survey, precluding a proper statistical analysis. In addition, the integrated brightness of  the system as well as that of their primaries is above the $K=12.5$ limit adopted in defining our sample. We note, however, that if these systems are on the close side of the cluster, as suggested by \citeauthor{siegler03}, they could be of equally low mass as some of our targets. Similarly, we do not address the multiplicity statistics of our subsample of low-mass cluster members since it is biased towards previously known or suspected binaries.

Despite its admittedly limited size, our sample is nonetheless complete for very low-mass stars in the Hyades, since we have observed all known cluster members with $K \geq 12.5$ discovered through large surveys of the cluster. It therefore provides a valuable counterpart to multiplicity surveys that targeted higher-mass members of the Hyades cluster \citep{reid97, patience98}. As the closest open cluster from the Sun, this population represents the second-best to probe multiplicity as a function of stellar mass after nearby field stars, prompting a comparison between the two populations. 

We have resolved 3 of the 16 very low-mass targets studied in this survey into close binaries, resulting in a raw companion frequency of 19$^{+13}_{-6}$\% (1$\sigma$ confidence intervals based on binomial statistics) over the 2--350\,AU separation range. This is similar to the companion frequencies for field very low-mass objects \citep{bouy03, burgasser07}, as well as for members of the Pleiades open cluster based on similarly high resolution imaging imaging \citep{martin03, bouy06}, although the latter were sensitive to slightly wider companions due to the larger distance to the cluster. In addition, photometric studies of substellar members of open clusters consistently derive multiplicity rates on the order of 25--30\% \citep[e.g.,][]{lodieu07, boudreault12}. The lower multiplicity rate derived here can be accounted for by systems with separations tighter than 2\,AU, whose frequency is on the order of 5--10\% among field stars \citep[][and references therein]{basri06, duchene13}.

\begin{figure}
   \centering
   \includegraphics[width=\hsize]{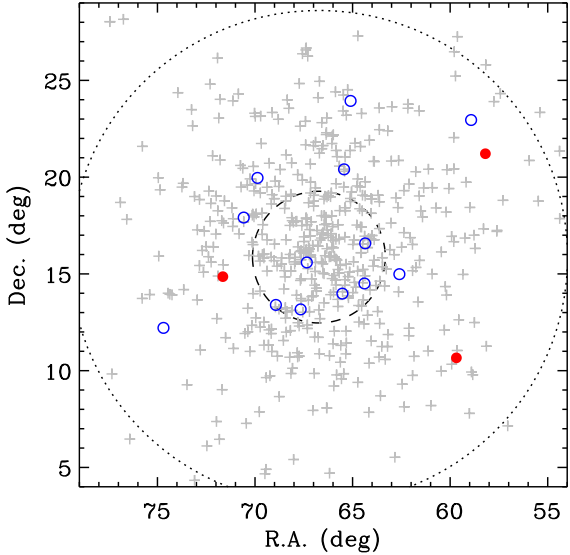}
      \caption{Spatial distribution of Hyades cluster members. Gray crosses represent probable members from the Prosser \& Stauffer database (J. Stauffer, priv. comm.), while open blue and filled red circles represent single and binary members in our sample of very low-mass members. The dashed and dotted circles represent the core and tidal radii from \cite{perryman98}, respectively.
              }
         \label{fig:cluster}
   \end{figure}

It has been suggested that the cluster has experienced severe mass segregation based on its apparently depleted population of substellar objects \citep{bouvier08}. However, we find no evidence that binary systems are over-represented among the substellar objects still present in the cluster. Indeed, the three binary systems discovered in this survey lie on the outskirts of the regions surveyed by \cite{hogan08}, and most apparently single stars lie closer to the cluster center (see Figure\,\ref{fig:cluster}). Similarly, the most massive single stars and binary systems in our survey are scattered somewhat more broadly from the cluster center than the lowest mass ones, albeit only marginally. This suggests that mass segregation occurs evenly to all systems with mass $\lesssim 0.1\,M_\odot$ irrespective of multiplicity, as expected from the perspective of cluster dynamical evolution. 

To allow a comparison of the frequency of visual binaries we find for substellar objects with those reported by \cite{reid97} for low-mass stars and \cite{patience98} for AFGK stars in the Hyades cluster, it is important to note that these studies probed different separation ranges. Indeed, the completeness ranges of the latter two surveys are 14--825\,AU and 5--50\,AU, respectively, although both studies also discovered tighter systems down to essentially the same limit as our own survey. Furthermore, each survey has a different minimum mass ratio limit \citep[e.g., $q \gtrsim 0.25$ for][]{patience98}, further blurring direct comparisons between surveys. Nonetheless, narrowing down to the 3--50\,AU separation range, probed by each surveys albeit possibly incompletely, we find remarkably similar companion frequencies across all masses: 17$^{+6}_{-4}$\% and $16^{+3}_{-2}$ for low-mass and AFGK stars, respectively. While this seems to be at odds with the overall decline of the multiplicity frequency towards lower primary masses, \cite{duchene13} pointed out that the companion frequency in the limited separation range 1--10\,AU appears constant across primary masses up to at least 1.5\,$M_\odot$ among field stars, in agreement with the result we obtain for the Hyades cluster. Ultimately, a more informative comparison awaits a search for spectroscopic binaries among Hyades very low-mass stars to complement the existing samples of such systems at higher masses \citep{bender08, reid00}.

\begin{figure}
   \centering
   \includegraphics[width=\hsize]{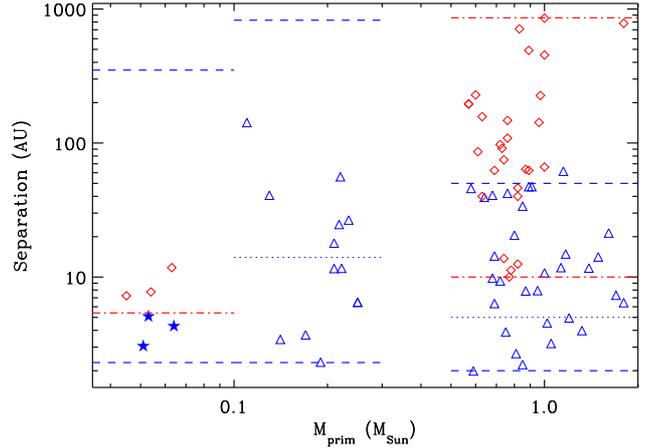}
      \caption{Projected separation of visual binaries as a function of primary mass for the Hyades (blue triangles and stars) and Pleiades (red diamonds) open clusters. The filled blue stars represent the very low-mass tight binaries discovered in this survey. The blue dashed lines marked the entire ranges of separation searched for companions in this survey, \cite{reid97} and \cite{patience98} in order of increasing primary mass while dotted lines indicate the completeness-defining minimum separation of the latter two surveys. The red dot-dashed lines indicate the equivalent limits for surveys in the Pleiades cluster \citep{bouvier97, bouy06}.
              }
         \label{fig:sep_mass}
   \end{figure}

An immediately apparent feature of the binaries we uncovered in this survey is their tight separation. As Figure\,\ref{fig:sep_mass} illustrates, all companions are clustered just beyond the minimum separation probed in our survey, in the 3--5\,AU range. On the other hand, for solar-type stars, \cite{patience98} discovered companions at all separations within their search region. Regarding low-mass stars in the Hyades, the survey of \cite{reid97} found a broad range of separations but an apparent deficit of companions outside of $\sim$150\,AU, despite being sensitive to companions up to 825\,AU. Binary systems in the Pleiades open cluster show a similar behavior \citep{bouvier97, martin03, bouy06}. Thus there appears to be a maximum binary separation that decreases sharply from solar-type stars to very low-mass stars in open clusters. This is reminiscent of findings among both field systems, where several empirical "maximum separation" laws have been proposed \citep{reid01, burgasser03, close03}, and star-forming regions \citep[e.g.,][]{kraus11}. In the very low-mass regime, this limit is around 25\,AU \citep[e.g.,][]{burgasser07}. Although a small fraction of "unusually wide" systems have later been identified \citep[e.g.,][]{bejar08}, their frequency is so low that we would not expect to detect any in our relatively small sample. Even among young stellar populations, where a handful of systems with separations up to $\sim$300\,AU have been found \citep[e.g.,][]{luhman09}, such systems remain rare considering the ease with which they can be discovered. The tight binaries we found in this survey therefore follow expectations from very low-mass binaries in both the field population and star-forming regions.

Given the tight separation of the systems we discovered here and the dynamically evolved state of the Hyades cluster, we expect that they will survive the remaining dissolution phase of the cluster and will thus be released in the Galactic field. While a large fraction of stars are believed to form in "clusters" \citep{lada03}, it is unclear whether rich open clusters, such as the Hyades, are truly representative of star formation as a whole \citep{bressert10}. Overall multiplicity statistics favor rich clusters as the birth place of the majority of field stars \citep[][and references therein]{duchene13}, in line with the fact that the multiplicity frequency we derive here for Hyades members matches that of field objects. However, a finer study that also takes all properties of binary systems into account (semi-major axis, mass ratio, eccentricity) is needed to confirm this conclusion. In the following, we consider the mass ratio distribution of very low-mass binaries.

\subsection{On the mass ratio distribution of very low-mass binaries}
\label{subsec:massratio}

\subsubsection{Comparison to systems in the field and in star-forming regions}

The most intriguing property of the Hyades very low-mass binary systems we have found is arguably their mass ratio, which are in the 0.69--0.84 range based on their $K$ band flux ratio. As shown in Figure\,\ref{fig:qlim2}, this prevalence of "intermediate" mass ratios was also observed in the Pleiades by \cite{martin03}, who used visible wavelength photometry to estimate binary mass ratios, however. While these authors ascribed it to small number statistics, the fact that we find the same trend in the Hyades instead suggests that this is common characteristics of open cluster brown dwarf binaries in general. In comparison, the overall population of binary field brown dwarfs is heavily skewed towards systems with $q \gtrsim 0.85$ \citep[e.g.,][]{burgasser07, duchene13}. Comparing the distribution of mass ratios between field\footnote{This is done using the Very Low-Mass Binary archive (http://vlmbinaries.org/). Although more systems have since then been discovered, they do not appear to skew the mass ratio distribution dramatically.} and open cluster very low-mass binaries using a Wilcoxon test yields a probability of $>99.8$\% that the two parent populations are different. We further note that the mass ratios of the two binaries from \cite{siegler03}, which are close in mass to the very low-mass systems considered here but are not included in this analysis, are also in the 0.75--0.85 range. In addition, using the mass ratios estimated from shorter wavelength observations for our Hyades binaries would further increase this discrepancy as they are systematically lower than those derived from the $K$ band brightness. Therefore, this apparent difference in mass ratio distribution between open cluster and field very low-mass binaries is highly significant.

Interestingly, the mass ratio distribution of very low-mass objects in star-forming region\footnote{We include here all systems with $M_{prim} \lesssim 0.1 M_\odot$ and with an age $\leq 5$\,Myr from the Very Low-Mass Binary archive as well as additional systems from \cite{kraus12}.} also appears to be much flatter than that of field objects (see Figure\,\ref{fig:qlim2}). Although larger uncertainties are likely associated with mass estimates in the youngest systems, objects with $q \leq 0.8$ are relatively common and systems with $q \leq 0.5$ exist \citep[e.g.,][]{konopacky07, todorov10, kraus12}. Given our limited sensitivity to companions at short separations, the observed distribution of mass ratio in very low-mass open cluster members could well be consistent with an essentially flat intrinsic distribution extending to planetary-mass companions. Similarly, \cite{kraus12} performed a Bayesian analysis of all systems observed at the highest resolution and found a relatively shallow mass ratio distribution for primaries in the 0.07--0.15\,$M_\odot$ range. Even though these authors identified a tendency for tighter systems to have mass ratios closer to unity, their sample include at least three systems with $q \leq 0.8$ that are closer than 10\,AU, which is extremely hard to reconcile with the observed distribution for field binaries. Thus, while the mass ratio distribution of very low-mass binaries in star-forming regions and open clusters are broadly consistent with one another, at least down to $q \sim 0.7$, the field distribution differs from both. 

If the difference in mass ratio distribution between "young" ($\lesssim 1$\,Gyr) and field systems is real, one may wonder about its physical origin. As pointed out above, it is unclear whether most field objects form in rich clusters, which could naturally account for this difference. However, the same contradiction is found if one assumes that most stars form in less dense star-forming regions instead. A mixture of both channels is most likely to feed into the Galactic population, and it is unlikely that most star formation on the galactic scale occurs in environments that are neither clusters nor loose associations. Therefore, the discrepancy in mass ratio distribution must be accounted for by a different explanation. Since the binding energy and total energy of a system is only marginally smaller if $q  = 0.7$ instead of $q = 0.9$, it is difficult to imagine that dynamical effects can explain the observed difference. Indeed, dynamical evolution of a cluster does not alter significantly the mass ratio distribution of binary systems \citep{parker13}. In principle, the slightly enhanced metallicity of the Hyades population \citep{boesgaard90} could have an impact on either core fragmentation or the subsequent accretion of material during the first few Myr of evolution of binary systems. However, this does not apply to the Pleiades members, which have an essentially solar abundance of metals, casting doubts on this interpretation.

In the following, we explore an alternative interpretation, namely that the apparent difference in mass ratio distribution is not real but stems from other, hidden factors. In particular, we consider the following possibilities: systematic errors in model-based mass estimates, subtle selection biases, and the possibility of a missing population of lower mass ratio systems among field binaries.

\subsubsection{Systematic biases in model-derived masses}

The first possible explanation consists in a mis-estimate of individual masses, hence mass ratios, in either population. For instance, the mass ratios derived in young systems could be significantly biased towards lower values due to systematic errors in evolutionary models. The fact that the mass ratios estimated from the $H$ and $J$ photometry are even lower than those from the $K$ photometry for the Hyades binaries seems to indicate otherwise, but this cannot be entirely excluded. In particular, the two lowest mass ratio systems, UKIDSS\,5 and 9, contain secondaries with $K > 15$, where the BT-Settl, DUSTY and COND models appear to diverge from one another and from the empirical sequence defined by bona fide cluster members. Nonetheless, all three models yield very similar (surprisingly, to within 2-3\%) predicted mass ratios when using the components' $K$ band photometry, albeit with different dependencies on the observing wavelength of the derived mass ratios and absolute masses. The fact that different models have been used to estimate mass ratios of field systems could also introduce an additional bias, but, again, the observed differences in mass ratio between the various models seem to be relatively small.

Irrespective of the details of these three sets of models and of the absolute masses of the individual objects, none of the tight binary systems in the Hyades or Pleiades is close to being equal-flux (in the near-infrared and visible, respectively). In order for these systems to have mass ratios in line with field systems ($q \gtrsim 0.9$), the mass-luminosity would need to be much steeper than in current evolutionary models for substellar systems with an age of several 100\,Myr. While it is currently impossible to exclude this possibility, we consider it unlikely. Among the rare very low-mass binary systems for which individual component masses could be determined independently of models, GJ\,569\,B, PPl\,15 and LP\,349-25 all three have ages estimated to be $\lesssim 500$\,Myr \citep{liu10}. Their mass ratios are in the 0.5--0.85 range \citep{basri99, zapatero04, konopacky10} and their observed flux ratios are in the 0.3--1\,mag range at both visible and near-infrared wavelengths. This confirms that even moderately non-equal flux ratios translate into mass ratios significantly lower than unity in relatively young systems. We thus believe that the relatively large magnitude difference in our Hyades very low-mass binaries are exceedingly hard to reconcile with a mass ratio that is very close to unity.

Alternatively, it could be that field very low-mass binaries have mass ratios that are consistently over-estimated. This would occur if the mass-luminosity relationship predicted by models was in fact much shallower than models predict, although this seems to be refuted by the same argument as above. Another possible explanation could be that field systems have their ages systematically over-estimated. For instance, consider the mass ratio for the 2MASS J1209-1004 system estimated by \cite{liu10}: if the system was as old as 10\,Gyr, its mass ratio would be $q \approx 0.9$, but if its age is $\lesssim 1$\,Gyr, then $q \approx 0.5$. Thus is most field binary brown dwarfs are relatively young, their intrinsic mass ratio distribution could be much flatter than believed to date. However, it is unclear why system ages could be so severely and systematically over-estimated. In addition, one expects systems of all ages in the Galactic field, so that the absence or rarity of systems with ages of several Gyr would need to be explained.

In summary, while model-derived mass ratios are likely biased in both "young" (0.1--1\,Gyr) and old systems, we do not believe that systematic errors between both populations can be large enough to account for the observed difference in the inferred mass ratio distributions.

\subsubsection{Selection biases}

A second explanation would call for a selection bias against equal-mass binaries in open cluster and star-forming regions. As pointed out in Section\,\ref{sec:obs}, our Hyades sample may be biased against equal-mass systems with components that are just below the 0.1$M_\odot$ maximum mass considered here. However, we note that the mass ratio distribution of field binaries does not seem different for substellar and very low-mass stellar primaries (see Figure\,\ref{fig:qlim2}). Had we focused strictly on substellar systems, the Hyades and field mass ratio distributions would still be different and this selection bias would not apply. Indeed, equal-mass systems have similar colors and proper motion as single objects and, although they would be somewhat brighter than unequal mass systems, their brightness would not place them outside of typical selection criteria given the broad sequence of known cluster members (see Figure\,\ref{fig:cmd}). Similar arguments apply to star-forming regions, where both kinematic and photometric criteria are even less stringent. For instance, a factor of 2 in total luminosity, typical of an equal-mass binary, is not nearly as large as the observed luminosity spread in a given population, so that it would be impossible to rule out membership on this sole criterion. We cannot think of any reason why equal-mass very low-mass binaries would have been missed in surveys of star-forming regions. We thus conclude that the difference we observe does not stem from selection biases in young open clusters or star-forming regions.

\subsubsection{Missing binaries among field very low-mass objects}

Arguably the most plausible explanation for the difference in mass ratio distribution is that the frequency of systems with $q \lesssim 0.8$ has been severely underestimated in the field population, where unequal mass systems are characterized by much steeper contrast ratios than among younger systems. In this context, it is worth noting that \cite{allen07} derived a much shallower power law index for the mass ratio distribution of field very low-mass binaries than other estimates \citep[e.g.,][]{burgasser06, liu10}, which would predict a significant number of systems in the $0.6 \leq q \leq 0.8$ range to which the binaries we discovered in the Hyades belong. The very low-mass binaries found in open clusters therefore suggest that a significant population of "intermediate" mass ratios still escapes detection among field stars as a result of the challenge currently posed by the detection of high-contrast, tight binaries. In turn, this would imply that the overall multiplicity fraction of field very low-mass stars has been significantly under-estimated by current studies. Recent discoveries of high-contrast and low mass ratio systems among field very low-mass objects provide additional circumstantial evidence that this is indeed the case \citep{pope13,sahlmann13}.

A much shallower mass ratio distribution for substellar binaries would in turn have consequences for theories that address the formation and evolution of such systems. Although the debate is not fully settled, the broad consensus is that substellar binaries form from the fragmentation of the lowest mass prestellar cores in the same way as stellar binaries do \citep{luhman12}. The marked prevalence of near equal-mass systems would require that the initial fragments are of almost equal mass, or that subsequent accretion systematically favors the least massive fragment to bring the final mass ratio of the system as close to unity as possible. Taken to the extreme, either scenario requires an uncomfortably fine-tuning of the physics at play. A significant population of lower mass ratio systems, as suggested by our results, relaxes these tight constraints. Still, given the low mass of the parent core, the potential for accretion after the initial fragmentation is much smaller than for higher mass systems, so that the final mass ratio should remain closer to the initial one. In most simulations of core fragmentation and cluster formation, the fragments have commensurable mass \citep{delgado04, goodwin04, bate12}, which is directly related to the Jeans mass. In cases where subsequent accretion is limited, one thus expects systems to maintain mass ratios $\gtrsim 0.5$, in broad agreement with the observed distributions in open clusters and star-forming regions.


\section{Conclusions}

Using the Keck\,II adaptive optics system, we have conducted the first multiplicity survey of very low-mass stars and brown dwarfs in the Hyades open cluster. From a complete sample of 16 targets, we identify 3 previously unknown tight binaries, with separations in the range 0\farcs066--0\farcs11, which we believe are physically bound to their primaries. A number of faint, distant objects were also found. In most cases, astrometric monitoring and/or near-infrared colors exclude cluster membership. By extension, we believe that all candidate wide binaries are non-physical. The multiplicity frequency, 19$^{+13}_{-6}$\% over the 2--350\,AU range, and preference for tight ($\lesssim 5$\,AU) systems are consistent with what is observed in the field. When coupled with multiplicity surveys for low-mass and solar-type stars, our results also show the same trends as observed among field stars. When combined with observations of objects in the Pleiades, we find that open cluster very low-mass binaries, as well as those found in star-forming regions, have a mass ratio distribution that is skewed toward significantly lower values than those of field binaries, which heavily favor near-equal-mass systems. Although model-based mass estimates are subject to caution, we find it unlikely that systematic errors in the predicted mass-luminosity relationship can account for this difference. We also exclude that selection effects precluded equal-mass binaries to be found by large scale searches for very low-mass cluster members. Since the field population comes primarily from open clusters and it is unlikely that dynamical effects are responsible for the observed trend, we suggest that the frequency of systems with $q \lesssim 0.8$ has been underestimated in previous surveys, an hypothesis that can be tested with future high-contrast, high-resolution imaging of field brown dwarfs.

\begin{acknowledgements}
The authors are grateful to John Stauffer for providing us with his list of probable cluster members and to the referee, Simon Goodwin, for his rapid review of our manuscript. This research has been conducted in the framework of the ANR 2010 JCJC 0501-1“DESC”. The authors thank the observing assistants Carolyn Parker and Gary Puniwai, and support astronomers Marc Kassis and Jim Lyke for their help in obtaining the observations. This research has made use of the Simbad and Aladin tools, CDS, Strasbourg France, of data obtained as part of the UKIRT Infrared Deep Sky Survey, and of data products from the SDSS-III project, which has been funded by the Alfred P. Sloan Foundation, the Participating Institutions, the National Science Foundation, and the U.S. Department of Energy Office of Science. The SDSS-III web site is http://www.sdss3.org/. The W.M. Keck Observatory is operated as a scientific partnership among the California Institute of Technology, the University of California and the National Aeronautics and Space Administration. The Observatory was made possible by the generous financial support of the W.M. Keck Foundation. The authors also wish to recognize and acknowledge the very significant cultural role and reverence that the summit of Mauna Kea has always had within the indigenous Hawaiian community. We are most fortunate to have the opportunity to conduct observations from this mountain.
\end{acknowledgements}

\end{document}